\documentclass[a4paper,fleqn]{cas-dc}

\usepackage[authoryear]{natbib}

\usepackage{aas}

\def\tsc#1{\csdef{#1}{\textsc{\lowercase{#1}}\xspace}}
\tsc{WGM}
\tsc{QE}
\tsc{EP}
\tsc{PMS}
\tsc{BEC}
\tsc{DE}

\newcommand{\ond}{Ond\v{r}ejov}

\newcommand{\T} {{\sc Tess}}
\newcommand{\ms}{M$_{\odot}$}
\newcommand{\oc}{$O\!-\!C$}
\newcommand{\rs}{R$_{\odot}$}

\begin{document}
\let\WriteBookmarks\relax
\def\floatpagepagefraction{1}
\def\textpagefraction{.001}

\shorttitle{TESS light curves and period changes in BB~Per}

\shortauthors{M. Wolf, P. Zasche, M. Mašek et al.}

\title [mode = title]{TESS light curves and period changes in low-mass eclipsing binary BB~Persei}                      



%

\author[1]{Marek Wolf}[orcid=0000-0002-4387-6358]
\cormark[1]
\ead{marek.wolf@mff.cuni.cz}

\credit{Methodology, Analysis of the data, Writing, review \& editing}

\affiliation[1]{organization={Astronomical Institute, Faculty of Mathematics and Physics, Charles University},
    addressline={V Hole\v{s}ovi\v{c}k{\'a}ch 2}, 
    city={Prague},
    postcode={180 00}, 
    country={Czech Republic}}   
\cortext[cor1]{Corresponding author}
\author[1]{Petr Zasche}[orcid=0000-0001-9383-7704]
\credit{Photomeric observations, Analysis of the data}

\author[2]{Miloslav Zejda}[orcid=0000-0001-6231-3350]
\credit{Photometric observation, Analysis of the data}

\affiliation[2]{organization={Institute of Theoretical Physics and Astrophysics, Masaryk
    University},
    addressline={Kotl\'a\v{r}sk\'a 2}, 
    city={Brno}, 
     postcode={611 37}, 
    country={Czech Republic}
    }

\author[3,4]{Martin Mašek}[orcid=0000-0002-0967-0006]
\credit{Photometric observations, Analysis of the data}

\affiliation[3]{organization={FZU - Institute of Physics of the Czech Academy of Sciences},
    addressline={Na Slovance 1999/2}, 
    city={Prague}, 
     postcode={182 21}, 
    country={Czech Republic}
    }        
\affiliation[4]{organization={Czech Astronomical Society, Variable Star and Exoplanet 
    Section},
    addressline={V\'ide\v{n}sk\'a~1056}, 
    city={Praha 4}, 
    postcode={142 00},
    country={Czech Republic}
    }
 
\author[5]{Andrej Mudray}[]
\credit{Photometric observations}

\affiliation[5]{organization={Private Observatory},
    addressline={Chloumky 31}, 
    city={Holovousy}, 
     postcode={508 01}, 
    country={Czech Republic}
    }       

\author[1,4,6]{Hana Kučáková}[orcid=0000-0002-1330-1318]
\credit{Photometric observations}
 

\affiliation[6]{organization={Research Centre for Theoretical Physics and Astrophysics, Institute of Physics, Silesian University},
    addressline={Bezru\v{c}ovo n\'am. 13}, 
    city={Opava}, 
     postcode={746 01}, 
    country={Czech Republic}
    }    

\affiliation[7]{organization={
Mt. Suhora Observatory, University of National Education Commission},
    addressline={Podchorazych 2}, 
    postcode={30 084},
    city={Cracow}, 
    country={Poland}
    }        

\author[7]{Waldemar Og\l oza}[orcid=0000-0002-6293-9940]
\credit{Photometric observations}

\author[1,8]{Jaroslav Merc}[orcid=0000-0001-6355-2468]
\credit{Photometric observations}

\author[1,9]{Jan Kára}[orcid=0000-0002-1012-7203]
\credit{Photometric observations}

\affiliation[8]{organization={
Instituto de Astrof\'isica de Canarias, Calle Vía Láctea, s/n}, 
postcode={E-38205},
city={La Laguna, Tenerife},
country={Spain}
    }

\affiliation[9]{organization={
Department of Physics and Astronomy, University of Texas Rio Grande Valley},
     city={Brownsville},
     postcode={TX 78520}, 
     country={USA}
     }

\author[1]{Vojtěch Dienstbier}[orcid=0009-0001-8183-8758]
\credit{Photometric observations}





\begin{abstract} 
We present a detailed analysis of the low-mass detached eclipsing binary
system BB~Persei, which contains two K-type stars in a circular orbit with a short period of 0.4856~d. 
We used light curves from the Transiting Exoplanet Survey Satellite (\T), which observed BB~Per in five sectors, to determine its photometric properties and a precise orbital ephemeris.
The solution of the \T\ light curve in {\sc Phoebe} results in a detached configuration, where the temperature of the primary component was fixed to $T_1 = 5\,300$~K according to {\sc Lamost}, which gives us $T_2 = 5\,050 \pm 50$~K for the secondary. 
The spectral type of the primary component was derived as K0 and the photometric mass ratio was estimated $q = 0.90 $. 
Slow period changes on the current O-C diagram spanning the past 25 years indicate the presence of a third body orbiting the eclipsing pair with an orbital period of about 
22~years.
The companion could be a red dwarf of spectral type M6 -- M7 with a minimal mass of about 0.1~\ms.
The characteristics and temporal variation of the dark region on the surface of the secondary component were estimated. 
\end{abstract}


\maketitle


\section{Introduction}

Despite low-mass stars ($M$ < 1 \ms) and their binary and multiple systems are the most common type of stars in our Galaxy, their origin and evolution are still not fully understood questions in star formation theory.
Several studies of low-mass binaries (LMB) found discrepancies between the results of observations and the theoretical structure and evolution models. The general conclusion was that the observed radius of low-mass stars in eclipsing binaries is $\sim 5-10$\% larger than predicted by models, while the effective temperatures are $\sim 10$\% cooler \citep{2002ApJ...567.1140T,  2003A&A...398..239R, 2006Ap&SS.304...89R}. 
The need of studying the LMB is growing as an important source of information about the possible existence of extrasolar planets. 
See also the current review by \cite{2022Galax..10...98M}.

Long-term photometric monitoring of late-type eclipsing binaries is also a very useful tool for studying star-spot parameters (their structure, coordinates, sizes and temperatures), their evolution and statistical properties.
The low-mass stars are also affected by chromospheric activity caused by a strong magnetic field, dark, or bright spots. This variable activity has been frequently observed as flares and plays an important role for precise determination of fundamental physical parameters, e.g. their radii and temperatures. Their rotation periods are synchronized with the orbital period as a result of the tidal forces.

Here we report on a long-term mid-eclipse-time campaign, light-curve modeling, and surface activity of a K-type eclipsing binary.
This system is a relatively known, but seldom investigated northern-hemisphere object, whose uninterrupted observations lasts almost 10~years.
The short orbital period about 12~hours, brightness up to 13~mag 
and the Declination over +30~deg makes it easily accessible object on the northern hemisphere for most of the year. 
To our knowledge, no precise photometric analysis nor period studies of BB~Per have been published so far. 

This paper is a continuation of our previous studies of period changes in LMB  \citep{2018A&A...620A..72W, 2021A&A...647A..65W}. 
The presented text is structured as follows. 
All available information about BB~Per is collected in Section~2.
In Section~3, we describe our photometric observations and data analysis.  
The new mid-eclipse times and \oc\ diagram are presented in Section~4. 
In Section~5, we analyze BB~Per light curves and derive its photometric parameters.  
The discussion of the results is given in Section~6 and our conclusions are briefly presented in Section~7.

\begin{table}
\begin{center}
\caption{{\sc Gaia} DR3 astrometric and photometric data on BB~Per.}
\label{tg}
\smallskip
\begin{tabular}{cl}
\hline\hline\noalign{\smallskip}
 Parameter              &  Value     \\
\noalign{\smallskip}\hline\noalign{\smallskip}
$\alpha_{2000}$ [h m s] & 03 55 25.61  \\
$\delta_{2000}$ [d m s] & +31 30 47.88 \\
pm $\alpha$ [mas/yr]    & 10.485   $\pm$ 0.017 \\
pm $\delta$ [mas/yr]    & --37.250 $\pm$ 0.013 \\
parallax [mas]          & 4.540  $\pm$ 0.016 \\
\noalign{\smallskip}\hline\noalign{\smallskip}
$B$ [mag]                 & 14.29    \\
$V$ [mag]                 & 12.658  $\pm$ 0.04 \\
$R$ [mag]                 & 12.11    \\
$G$ [mag]                 & 12.238  $\pm$ 0.005 \\
$BP$ [mag]                & 12.930  $\pm$ 0.005 \\
$RP$ [mag]                & 11.403  $\pm$ 0.005 \\
$J$ [mag]                 & 10.835  $\pm$ 0.022 \\
$H$ [mag]                 & 10.285  $\pm$ 0.019 \\
$K$ [mag]                 & 10.120  $\pm$ 0.018 \\
\hline
\end{tabular}
\end{center}
\end{table}

\section{BB~Persei}

The low-mass eclipsing binary BB~Per 
(also SVS 557, 
TIC~94470667, 
NSVS 6705346, GSC 02357-00743, 	
$V_{\rm max} = 12.7$ mag) 
is a rather neglected northern object with a short orbital period of about 12~h.  
Its variability was discovered photographically by \cite{1969MmSAI..40..261R} 
in the field of o Per and classified as an irregular variable. 
Using the publicly available Northern Sky Variability Survey \citep [NSVS]{2004AJ....127.2436W} BB~Per was later reclassified as a probable low-mass eclipsing binary by \cite{2006IBVS.5674....1O} with the light ephemeris: 
$${\rm Pri.Min. = HJD\; 24\;51488.631 + 0.485594} \cdot E. $$

\noindent
They also mentioned the significant light changes of the binary: 12.56 -- 13.31 mag.
BB~Per is also listed as an object 2RXP J035525.1+313048 in the 
Second ROSAT PSPC Catalog \cite{2000yCat.9030....0R} as a faint X-ray source. 
The Catalina Surveys periodic variable stars \citep{2014ApJS..213....9D} give a similar period of 0.4856060 d and an amplitude of 0.49 mag.
BB~Per is also included in the catalog of eclipsing binaries observed by LAMOST
\citep{2018ApJS..235....5Q}, where the mean effective temperature 5365~K is given
and two values of radial velocities are listed: --17.0 and +10.4~km/s. 
BB~Per is also included in the ZTF catalog of periodic variable stars \cite{2020ApJS..249...18C}, where an improved period of 0.485608~d is presented.
The following linear ephemeris was proposed in the VSX index~\footnote{\url{https://www.aavso.org/vsx/}} for current use with a slightly modified period:
$$ {\rm Pri.Min. = HJD\; 24\;51488.631 +0.48560845 } \cdot E. $$

\noindent
The {\sc Gaia} DR3 astrometric and phometric data on BB~Per are summarized in Table~\ref{tg} \citep{2022yCat.1355....0G}. 
The distance to the system was derived to be $d \simeq 220$ pc \citep{2021AJ....161..147B}.

\section{Photometric observations}

In this study, we used our new ground-based photometric measurements
as well as archival photometry accessible from the \T\ satellite. 

\subsection{Ground based photometry}

Since 2012 the time-resolved CCD photometry of BB~Per, mostly during eclipses, has been regularly obtained at several observatories. Smaller telescopes and the modern CCD technique were sufficient for good S/N photometry of this relatively bright object.

\begin{itemize}

\item \ond\ observatory, Czech Republic: Mayer 0.65-m ($f/3.6$) reflecting telescope, CCD camera G2-3200 and photometric BVRI filters. Exposure times of
30 -- 60 seconds were used.

\item Masaryk University Observatory, Czech Republic:
Newtonian 0.60-m telescope, CCD camera G4-16000 and Johnson's VRI filters at Brno, or AZ~800 telescope, CMOS camera C5 and Sloan r, g photometric filters at Ždánice station. Exposure times of 60 sec were used on both telescopes. 

\item  Adiyaman Observatory, Turkey,
0.60-m photometric telescope, CCD camera Andor iKon-M 934 and Johnson's R filter.
Individual exposure times lasted 60 seconds. 

\item  Private observatory of A.M. in Holovousy near Ho\v{r}ice, Czech Republic, the SCT 0.406-m ($f/6.3$) telescope, reducer and coma corrector Starizona 0.63x, CMOS camera ZWO ASI071. The clear filter and the 60-sec exposure time were used.

\end{itemize}

\noindent
Our CCD observations at all observatories were reduced in a standard way. The images were bias-corrected and flat-fielded before aperture photometry was carried out.
The {\sc C-Munipack}\footnote{Package of software utilities 
for reducing astronomy CCD images, current version 2.1.37, 
available at \url{http://c-munipack.sourceforge.net/}}, a synthetic aperture photometry software, was routinely used.
Time series were constructed by computing the magnitude difference between the variable and a nearby comparison and check star. Lastly, the heliocentric correction was applied. 
The computers at the telescopes are synchronized by using different time servers. 
These corrections are usually on the order of $10^{-3}$ seconds.

\subsection{\T\ photometry}

The precise monitoring of light curves is also possible from space, and their exceptional quality is perfect for modeling the system and studying eclipse timings.
As a northern object with a moderate declination, BB~Per was also frequently measured by the {\it Transiting Exoplanet Survey Satellite} 
\citep[\T,][]{2015JATIS...1a4003R}, in several sectors; see Table~\ref{tess}.
We collected simple aperture photometry (SAP) of available short cadence produced by the Science Process Operation Centre \citep{2016SPIE.9913E..3EJ} and available at the Mikulski Archive for Space Telescope (MAST)\footnote{ \url{https://mast.stsci.edu/portal/Mashup/Clients/Mast/Portal.html }}. 

New times of primary and secondary minima and their uncertainties were generally determined by fitting the light curve by Gaussians or polynomials of the third or fourth order. The least squares method was used. They are listed in Table~\ref{bbmin}, where the epochs were computed according to the following improved linear ephemeris:
\begin{equation}
{\rm Pri. Min. = HJD\; 24\; 51488.6365(1) + 0.485608503(5)} \cdot E. 
\label{ephem} 
\end{equation}

\begin{table}
\caption {\T\ visibility of BB~Per and sectors used for light curve analyses and for mid-eclipse time determination.}
\label{tess}
\smallskip
\begin{tabular}{cccccc}
\hline\hline\noalign{\smallskip}
   \T\      &  Start date   &  Exposure    \\
Sector No.  &  [YYYY-MM-DD] &  time [sec]  \\
\hline\noalign{\smallskip}
     18     & 2019-11-03  &  1800 \\
     43     & 2021-09-16  &  600  \\
     44     & 2021-10-12  &  600  \\
     70     & 2023-09-20  &  200  \\ 
     71     & 2023-10-16  &  200  \\
\hline
\end{tabular}
\end{table}

\noindent
The mid-eclipse times obtained in \ond\ and MUO were calculated as the mean value of the measurements in $VRI$ or $VR$ filters.

\begin{table}[t]
\begin{center}
\footnotesize 
\caption{Ground-based times of primary and secondary eclipses of BB~Per.}
\label{bbmin}
\begin{tabular}{lrlc}
\hline\hline\noalign{\smallskip}
BJD --     & Epoch & Error   &  Observatory    \\
24~00000   &       & [day]   &  Source   \\
\hline\noalign{\smallskip}
51397.8278   & -187.0 & 0.002 & VarAstro \\
51488.6318   &    0.0 & 0.002 & \cite{2006IBVS.5674....1O} \\
54416.37240  & 6029.0& 0.0001 & AAVSO \\
54445.50898  & 6089.0& 0.0001 & AAVSO \\
54446.23612  & 6090.5& 0.0001 & AAVSO \\
54446.48040  & 6091.0& 0.0001 & AAVSO \\
54454.49082  & 6107.5& 0.0003 & AAVSO \\
54460.31807  & 6119.5& 0.0002 & AAVSO \\
54471.48753  & 6142.5& 0.0001 & AAVSO \\
54516.4079   & 6235.0& 0.0004 & \cite{2010IBVS.5918....1H} \\
55808.6095   & 8896.0& 0.005  & \cite{2013OEJV..160....1H}\\
55966.43153* & 9221.0& 0.0001 & \ond \\
56203.65214* & 9709.5& 0.0001 & \ond \\
57367.4119  & 12106.0& 0.0003 &\cite{2017OEJV..179....1J} \\
57681.84167 & 12753.5& 0.0002 & AAVSO \\
58391.55873 & 14215.0& 0.0001 & MUO \\
58551.32412 & 14544.0& 0.0002 & MUO\\
58737.55403 & 14927.5& 0.0001 & MUO \\
58772.51783 & 14999.5& 0.0001 & MUO \\
58924.2709  & 15312.0& 0.0005 & AAVSO \\
59125.55639* & 15726.5& 0.0001 & \ond \\
59144.49547* & 15765.5& 0.0003 & MUO \\
59185.52909* & 15850.0& 0.0003 & MUO \\
59623.30567* & 16751.5& 0.0002 & MUO \\
60267.71051  & 18078.5& 0.0001 & AAVSO \\
60329.38213* & 18205.5& 0.0001 & MUO \\
60334.23800* & 18215.5& 0.0001 & \ond \\
60338.36574  & 18224.0& 0.0001 & MUO \\
60346.37864  & 18240.5& 0.0003 & Holovousy \\
60354.39125  & 18257.0& 0.0001 & Holovousy \\
60359.24761  & 18267.0& 0.0001 & MUO \\
60392.26821  & 18335.0& 0.0001 & Adiyaman \\
60526.53908  & 18611.5& 0.0002 & MUO \\
60549.60634  & 18659.0& 0.0001 & MUO \\
60590.39757  & 18743.0& 0.0001 & Holovousy \\
60600.83719  & 18764.5& 0.0002 & AAVSO \\
60607.39346  & 18778.0& 0.0002 & Holovousy \\
60645.27097  & 18856.0& 0.0001 & MUO \\ 
60671.49277  & 18910.0& 0.0003 & MUO \\ 
\hline
\end{tabular}
\smallskip

Note: * mean value of VR, VI or VRI measurements
\end{center}
\end{table}

\begin{table}[t]
\begin{center}
\footnotesize 
\caption{New \T\ times of primary and secondary eclipses of BB~Per.}
\label{bbmintess}
\begin{tabular}{cccc}
\hline\hline\noalign{\smallskip}
BJD --     & Epoch &  BJD --     & Epoch   \\
24 00000   &       &  24 00000   &       \\
\hline\noalign{\smallskip}
\multicolumn{2}{c}{Sector 43} & \multicolumn{2}{c}{Sector 44}  \\
59474.22240 & 16444.5  &   59500.44513 & 16498.5       \\
59474.46654 & 16445.0  &   59500.68914 & 16499.0       \\
59474.70804 & 16445.5  &   59508.45862 & 16515.0       \\
59479.56405 & 16455.5  &   59511.61427 & 16521.5       \\
59479.80796 & 16456.0  &   59513.80036 & 16526.0       \\
59480.04965 & 16456.5  &   59517.68517 & 16534.0       \\
59484.90565 & 16466.5  &   59520.59882 & 16540.0       \\
59498.26109 & 16494.0  &   59521.32633 & 16541.5       \\
59498.50274 & 16494.5  &   59523.02675 & 16545.0       \\
59498.74669 & 16495.0  &   59523.99789 & 16547.0       \\
\noalign{\smallskip}\hline\noalign{\smallskip}
\multicolumn{2}{c}{Sector 70} & \multicolumn{2}{c}{Sector 71}   \\
60208.70782 & 17957.0   &  60234.93041 & 18011.0       \\
60212.83533 & 17965.5   &  60237.84429 & 18017.0       \\
60216.72030 & 17973.5   &  60241.24355 & 18024.0       \\
60225.21838 & 17991.0   &  60245.85723 & 18033.5       \\
60225.46131 & 17991.5   &  60250.95572 & 18044.0       \\
60228.61768 & 17998.0   &  60255.56938 & 18053.5       \\
60230.80303 & 18002.5   &  60256.29747 & 18055.0       \\
60231.28874 & 18003.5   &  60256.54059 & 18055.5       \\
60233.23111 & 18007.5   &  60257.02633 & 18056.5       \\
60233.47372 & 18008.0   &  60257.26871 & 18057.0       \\
\hline
\end{tabular}
\end{center}
\end{table}

\section{Orbital period changes}

\begin{figure*}[t]
\centering
\includegraphics[width=0.65\textwidth]{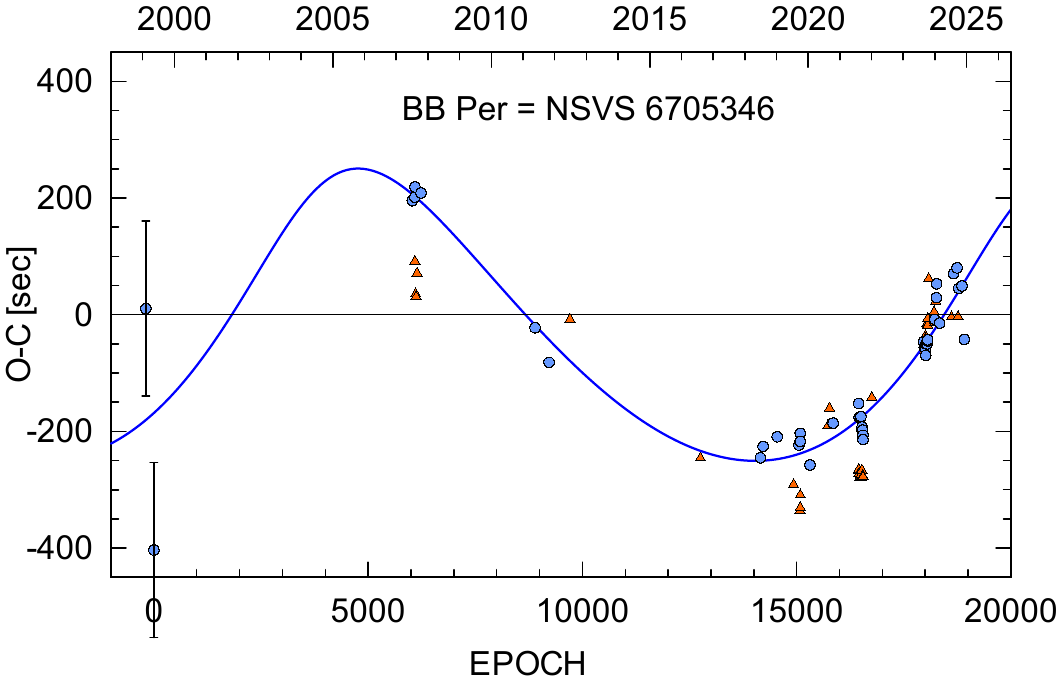}
\caption[ ]{The current \oc\ diagram for the times of minimum of 
BB~Per since discovery. The blue sinusoidal curve represents the LITE with a period of about 22~years and a semi-amplitude of about 250~sec. The individual primary and secondary minima are denoted by blue circles and orange triangles, resp. Due to surface activity the secondary minima are slightly shifted. 
\T\ minima are groups of points around the epoch 16~500 and 18~000.
The third-body orbital period is limited by large uncertainty of two initial mid-eclipse times.}
\label{bboc}
\end{figure*}


The period changes of BB~Per had not been studied since its discovery.
Many LMBs display periodic eclipse time variations caused by a third 
component orbiting the eclipsing pair (the so-called {\it light-time effect}, {\sc Lite}).
The {\sc Lite} in BB~Per were studied traditionally by means of an \oc\ diagram 
(or ETV curve) analysis.
As has been proven many times in the past, this simple and effective technique
is a very powerful tool for investigating the multiplicity of stellar systems.
For a detailed description of the LITE analyses, see the original papers 
of \cite{1952ApJ...116..211I, 1959AJ.....64..149I, 1990BAICz..41..231M, 
 2005ASPC..335.....S}.

Only a few eclipse timings have been reported in the literature
or is collected in VarAstro~\footnote{\url{https://var.astro.cz/en}}. 
We take into account all times given in Table~\ref{bbmin}, 
where 10 new mid-eclipse times were derived using the 
AAVSO~\footnote{\url{https://www.aavso.org/LCGv2/}} data.
Using the \T\ data, we calculated 40 very precise times
to complete our current \oc\ diagram. These times divided into four groups 
according to different \T\ sectors are included in Table~\ref{bbmintess}.
Their uncertainty was fixed to 0.00005~day although the least squares fit of the light curve during the eclipse gives us formally smaller values. However, the scatter of the \T\ minima is surprisingly larger than the given error. 
Because \T\ data are provided in the Barycentric Julian Date Dynamical Time (BJD$_{\rm TDB}$), all our times in Table~\ref{bbmin} were first transformed to this time scale using the generally used Time Utilities of Ohio State University\footnote{\url{http://astroutils.astronomy.ohio-state.edu/time/} }  \citep{2010PASP..122..935E}. 

A total of 88 mid-eclipse times, including 40 secondary eclipses, were used for the period analysis. The computed linear light elements and their internal errors of the least-squares fit are given in Eq.~\ref{ephem}.
Other parameters for the third-body orbit in BB~Per are listed in Table~\ref{t2}, 
where $P_3$ is the orbital period of the third body, $e_3$ is the eccentricity,
$A$ is the semi-amplitude of the light-time curve, 
$\omega_3$ the longitude of periastron and $T_3$ time of periastron.
The \oc\ diagram for the all mid-eclipse times is plotted in Fig.~\ref{bboc}. 
In addition, shallower secondary eclipses show rather larger deviations of their \oc\ values because they are probably more sensitive to some surface inhomogeneities of the secondary component. This effect was further studied by a light-curve solution.
The observable difference in the mid-eclipse times of spotted components was also described by \cite{2017AJ....154...30K}.

\begin{table}[t]
\caption{LITE parameters for BB~Per and the minimal mass of the third body
         (with errors of the last digit in parentheses). }
\label{t2}
\centering
\begin{tabular}{cccc}
\hline\hline\noalign{\smallskip}
Element & Unit      & Value       \\  
\noalign{\smallskip}\hline\noalign{\smallskip}
$T_0$ & BJD-2400000 & 51488.6365(1)      \\
$P_s$ & day         & 0.485608503(5)     \\
$P_3$ & day         & 8070(300)          \\
$P_3$ & year        & 22.1(0.8)          \\ 
$e_3$ &  --         & 0.33(2)            \\ 
$A$   & day         & 0.0029(2)          \\
$A$   & sec         & 250(20)            \\ 
$\omega_3$ & deg    & 65.2(2.0)          \\
$T_3$ &  JD-2400000 & 53290(50)          \\ 
$f(M_3)$         & \ms   &   0.000 27    \\
$M_{\rm 3, min}$ & \ms   &   0.096       \\
$M_{\rm 3, min}$ & M$_{\rm Jup}$   &  100   \\        
$K_3$            & m/s    &   730         \\
$a_{12}\sin i$   & au     &   0.51            \\ 
$A_{\rm dyn,3}$  &  days  &   $10^{-6}$      \\   
\hline \noalign{\smallskip}
$\sum{w\ (O-C)^2}$ & day$^2$ &  1.7$\cdot10^{-5}$      \\ 
\noalign{\smallskip}\hline
\end{tabular}
\end{table}

In case of a coplanar orbit of the third component ($i_3 \simeq i$), we can obtain a lower limit for its mass $M_{3, \rm min}$. 
Assuming $M_1 + M_2 = 1.68$~\ms\ (see next chapter), the possible companion could 
be a red dwarf of spectral type M6 -- M7 
($M_{\rm bol} \simeq 12.5$ mag) with a minimal mass of about 0.1~\ms, which
is practically invisible near two brighter K stars ($M_{\rm bol} \simeq 5 - 6$~mag).
The amplitude of the dynamical contribution of the third body, 
$A_{\rm dyn}$, is given by \citep{2016MNRAS.455.4136B}:

$$ A_{\rm dyn} = \frac{1}{2\pi} \frac{M_3}{M_1+M_2+M_3} \frac{P_s^2}{P_3} 
             \, \left(1-e^2_3\right)^{-3/2} $$

\smallskip
\noindent
and is given in Table~\ref{t2}. The value of $A_{\rm dyn}$ is on the order of tenths of a second and is less than the precision of individual mid-eclipse time estimation.

\section{Light curve solution}
\begin{figure}
\centering
\includegraphics[width=0.45\textwidth]{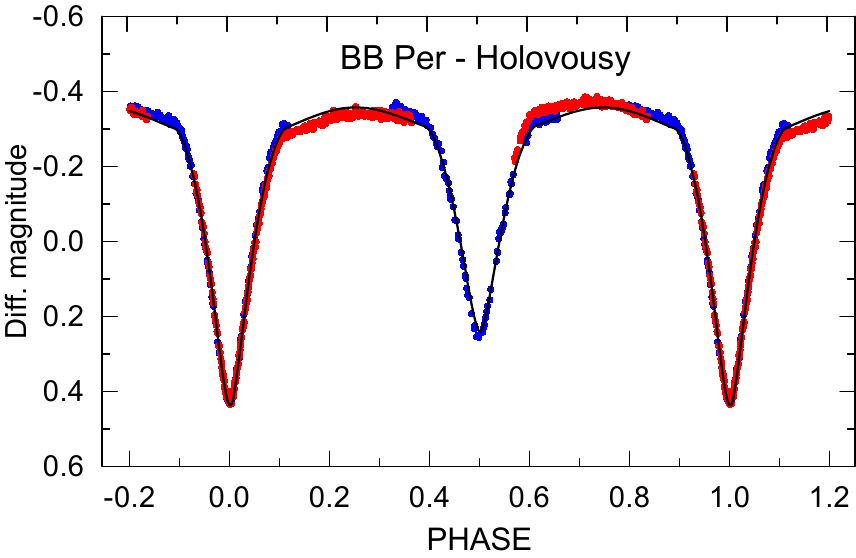}
\caption{Examples of our observations. The light curve of BB~Per obtained 
in a clear filter in Holovousy observatory near Hořice by A.M. in February 2024 (blue dots) and November 2024 (red points). Mean solution in {\sc Phoebe} is denoted as a black curve. The changes of the out-of-eclipse light curve 
are clearly visible. The nearby star UCAC4 608-010864 was used as 
a suitable comparison star. }
\label{BBMud}
\end{figure}

\begin{figure}[t]
\begin{center}
\includegraphics[width=0.45\textwidth]{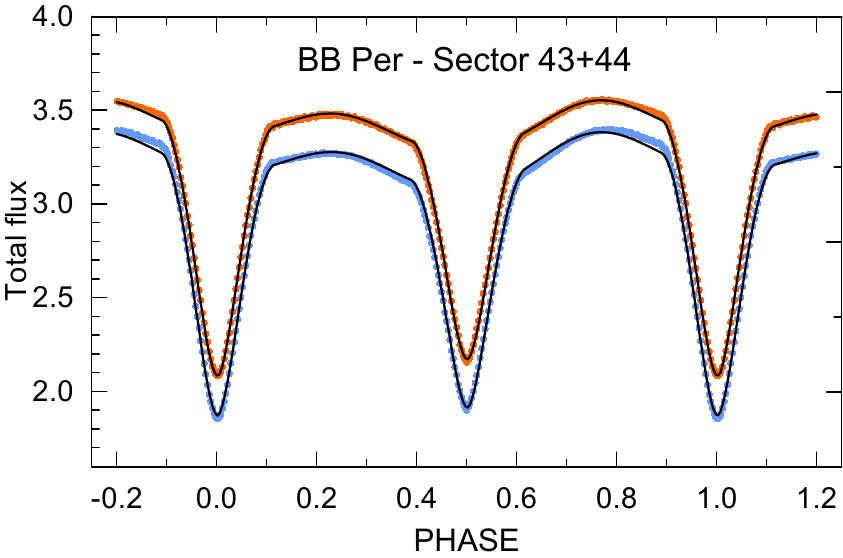}
\includegraphics[width=0.45\textwidth]{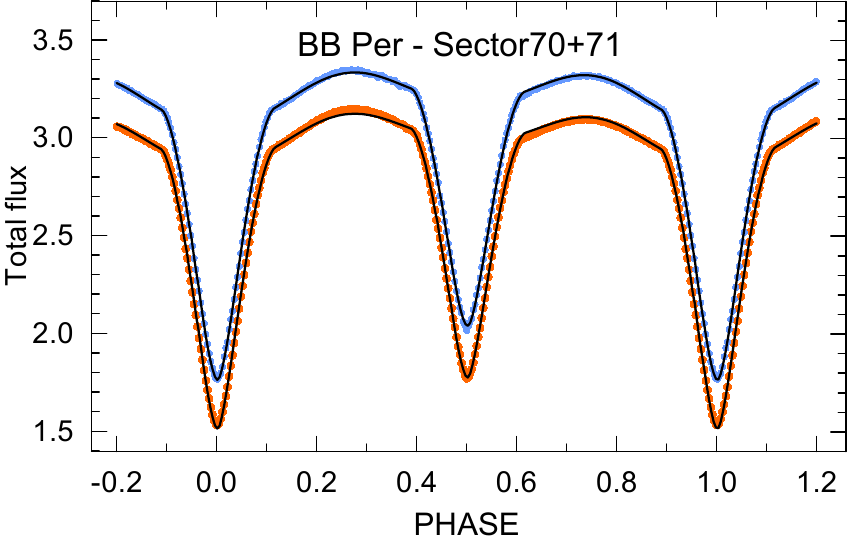}
\caption[ ]{The {\sc Phoebe} final solution for the \T\ light curves of BB~Per, upper panel: Sectors~43 and 44 (Sep/Oct 2021), lower panel: Sectors~70 and 71 (Sep/Oct 2023) - shifted for clarity. The \T\ data are denoted blue and orange, the resulting model in black. The change of brightness in maxima is clearly visible.} 
\label{BBTESS}
\end{center}
\end{figure}

\begin{figure}[t]
\begin{center}
\includegraphics[width=0.45\textwidth]{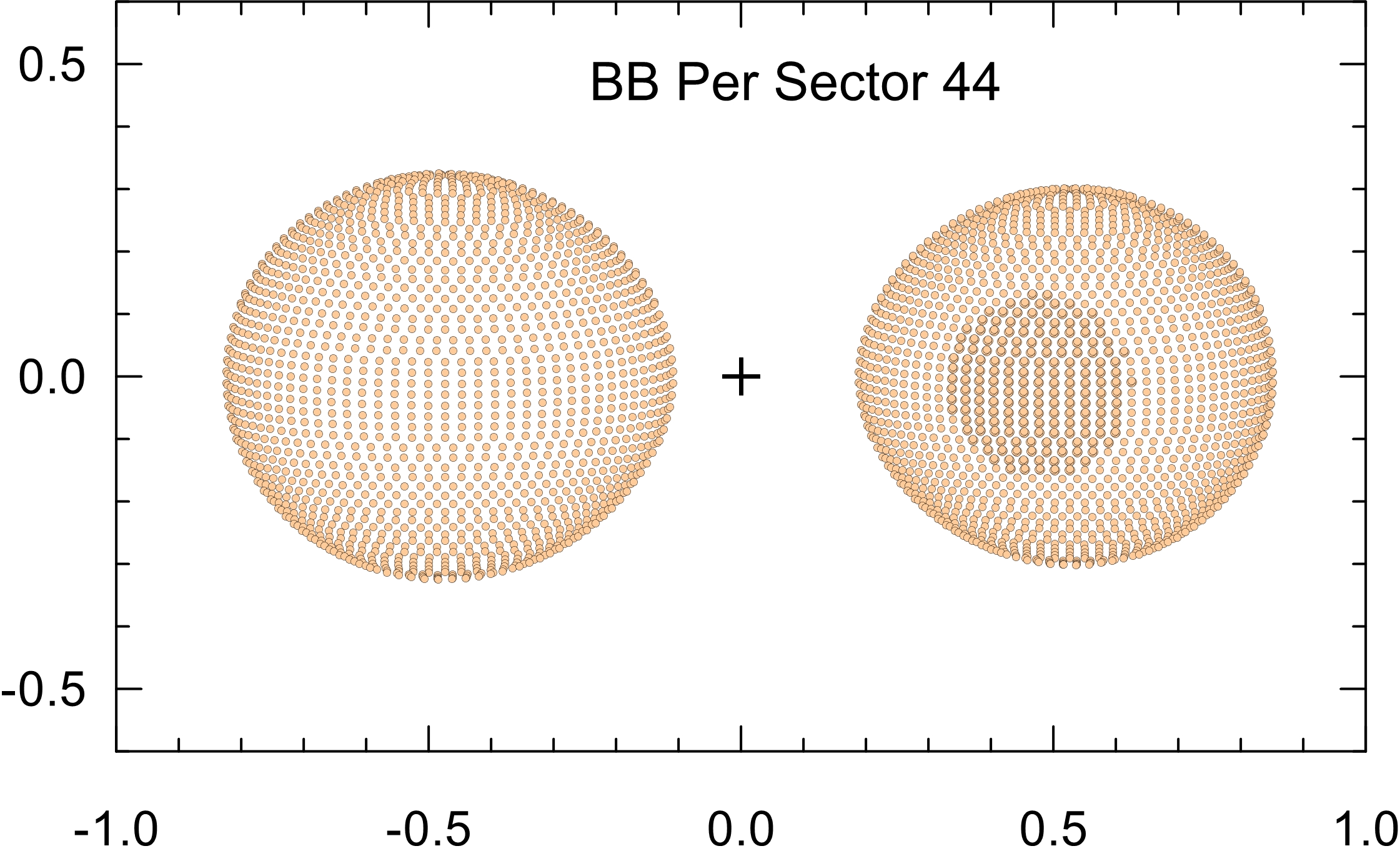}
\includegraphics[width=0.45\textwidth]{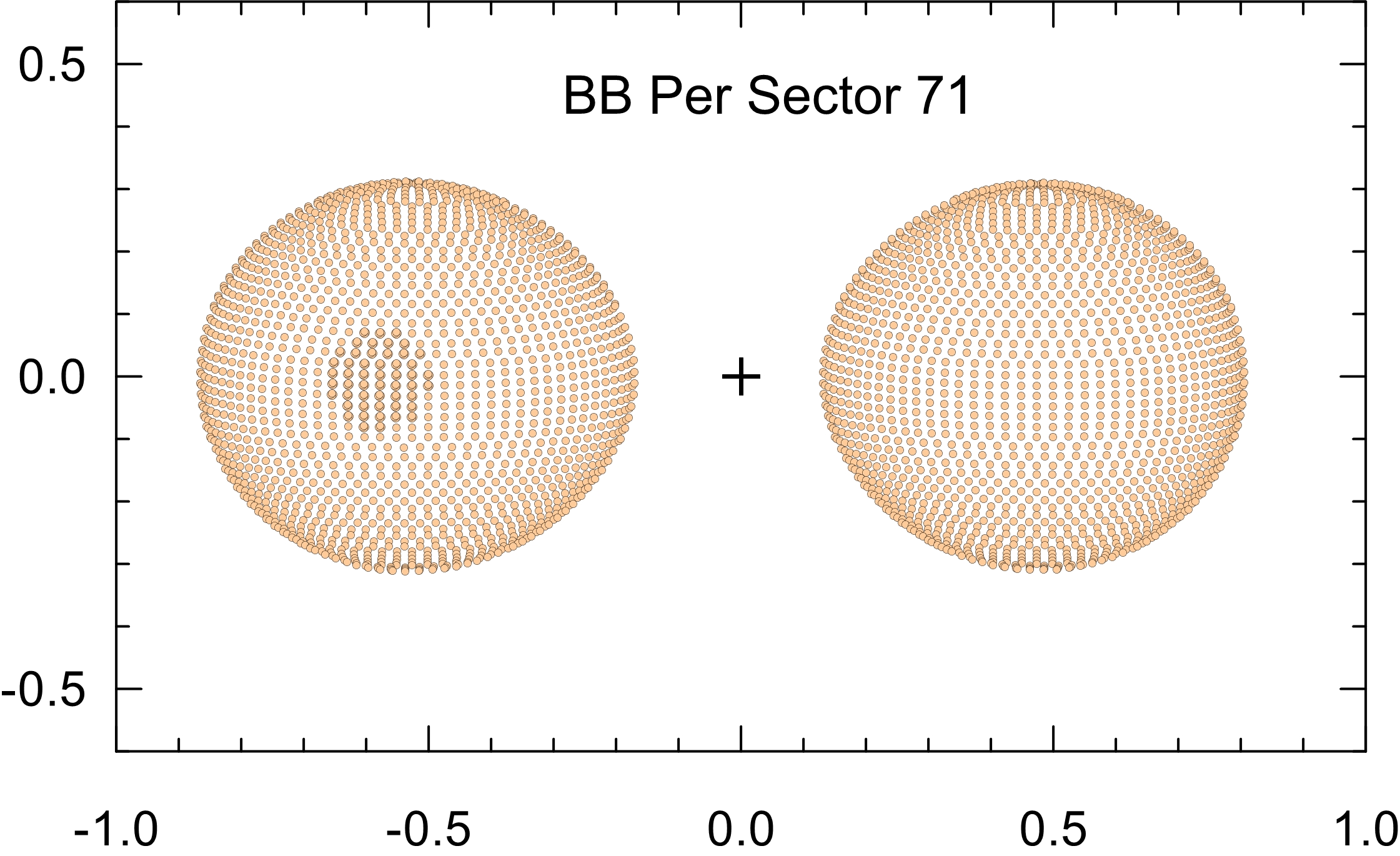}
\caption[ ]{3D models of BB~Per: upper panel in phase 0.25 for \T\ Sector~44, lower panel in phase 0.75 for \T\ Sector~71 with the dark structure on the secondary component. The cross marks the mass center of the system.}
\label{BB3D}
\end{center}
\end{figure}

To our knowledge, no photometric solutions have been reported for BB~Per so far. 
Comparison of \T\ light curves obtained in different epochs
clearly shows small changes of their shape, especially out of eclipse
around phases 0.25 and 0.75; see also example of our photometry in Fig.~\ref{BBMud}. 
All \T\ light curves of BB~Per with short exposure times obtained during 2021--2023 at our disposal were analyzed using the well-known code 
{\sc Phoebe}\footnote{\url{http://phoebe-project.org/}} \citep{2005ApJ...628..426P, 2018maeb.book.....P} which is based on the original Wilson-Devinney algorithm \citep{1971ApJ...166..605W} and is widely used to model the photometric light curves of the eclipsing binaries. The wavelength coverage of \T\ is about 6~000 -- 10~000~\AA, covering most of the R and I photometric bands.

In the literature a very wide temperature range is reported for BB~Per, 
from $T_{\rm eff}$ = 4444~K (Gaia DR2) to 5598~K \citep[Gaia DR3 and LAMOST,][]{2023ApJS..266...18Z}.
Thus, the temperature of the primary component was set at the mean value 
of 5~300~K given in the LAMOST archive. 
According to the table of \cite{2013ApJS..208....9P}
\footnote{\url{http://www.pas.rochester.edu/~emamajek/EEM\_dwarf\_UBVIJHK\_colors\_Teff.txt}}, this temperature corresponds to the spectral type K0 and a mass of 0.88 \ms. 
In the absence of the spectroscopic mass ratio, we used for a rough estimate of the mass ratio the following relation \citep{2003MNRAS.342.1334G, 2017AJ....154...30K}: 
\begin{equation}
\log q =  0.207 \, (\log L_{\rm 2,bol} - \log L_{\rm 1,bol}),
\end{equation}
\noindent where $L_{\rm 1,bol}$ and $L_{\rm 2,bol}$ are bolometric luminosities. We set the initial mass ratio $q = 0.95$ for two similar components and this value was corrected several times during the light curve solutions in the four \T\ sectors.
The frequently used $q$-search method was not applied because it is suitable primarily for semidetached or overcontact binaries only; in case of partial eclipses, they do not contain all the necessary information \citep{2022Galax..10....8T}.  

Because components of BB~Per belong rather to late-type stars, we adopted bolometric albedos and gravity darkening coefficients as $A_1=A_2=0.5$ and $g_1=g_2=0.32$, which correspond to convective envelopes. Synchronous rotation for both components of the system ($F_1=F_2=1$) and a circular orbit ($e=0$) was assumed. We used the logarithmic limb-darkening law with the coefficients adopted from \cite{1993AJ....106.2096V} tables.

At first we fitted those parameters that had a substantial effect on the shape of the light curve.
The adjustable parameters were thus the inclination $i$, the effective temperature of the secondary component $T_2$, the luminosity and dimensionless potentials of both components $\Omega_1, \Omega_2$. Later, the characteristics of a cold spot/region on the secondary component (colatitude, longitude, spot radius, and temperature factor) were then included. 
The need to include spots to the final solution was evident in view of the modulation of out-of-eclipse light curves. Usually, one or two spots on the surface of each component are used to fit the major asymmetries in the light curve. The high quality \T\ data supports this possibility. 
Because the effective temperature and the radius of a spot are strongly correlated, we assumed that the ratio of spot/star temperatures is close to 0.95 in our analyzes.

Implementation of the third light $L_3$ did not improve the fit, so we set $L_3 = 0$. 
The fine and coarse grid raster for both components was set to 30.
Binning 300 was used for all \T\ light curves obtained in one sector.
Numerous {\sc Phoebe} runs in detached mode using the different setup of initial parameters were evaluated. The parameters and the value of the cost function were recorded. 
The final photometric elements of the light curves are given in Table~\ref{t3},
where also the bolometric limb-darkening coefficients and the relative radii of both components are given.  The solution of \T\ light curves is plotted in Fig.\ref{BBTESS}.
As one can see, the agreement between the theoretical and observed light curves is very good. 
The resulting parameters of the cold region on the primary are given in Table~\ref{t8}. 
The geometrical representation of BB~Per in two epochs is shown in Fig.~\ref{BB3D}.

\begin{table}
\begin{center}
\caption{Adopted photometric elements of BB~Per, the mean values of {\sc Phoebe} solution in four \T\ sectors. }
\label{t3}
\begin{tabular}{ccccc}
\hline\hline\noalign{\smallskip}
            & 21.10.24 &  \\  
Parameter   & Primary  & Secondary \\
\noalign{\smallskip}\hline\noalign{\smallskip}
$i$ [deg]       &  \multicolumn{2}{c}{83.6(0.5)}   \\
$q$             &  \multicolumn{2}{c}{0.90 (fixed)  }  \\
$T_{1,2}$  [K]  &  5300 (fixed) &  5050 (50)    \\
$\Omega_{1,2}$  &  3.898      &  4.117      \\
$X_{1,2}$       &  0.638      &  0.648      \\
$r(\rm pole)$   &  0.322      &  0.290      \\
$r(\rm side)$   &  0.333      &  0.298      \\
$r(\rm point)$  &  0.363      &  0.320      \\
$r(\rm back)$   &  0.349      &  0.311      \\
\noalign{\smallskip}\hline
\end{tabular}
\end{center}
\end{table}

\begin{table}
\begin{center}
\caption{Parameters of the dark surface structure on the secondary
 component of BB~Per. }
\smallskip
\label{t8}
\begin{tabular}{lcccccc}
\hline\hline\noalign{\smallskip}
Parameter              & \multicolumn{4}{c}{\T\ Sector}   \\     
                       & 43   & 44    & 70    & 71  \\
\noalign{\smallskip}\hline\noalign{\smallskip}
Colatitude [deg]       & 70   & 75    & 80    &  80  \\
Longitude [deg]        & 250  & 250   & 118   &  118 \\
Spot radius [deg]      & 32   & 32    & 13    &  16  \\
Temperature factor     & 0.93 & 0.95  &  0.93 &  0.93  \\
{\sc Phoebe} cost function & 1050 & 550  & 1470 & 1210 \\      
\noalign{\smallskip}\hline
\end{tabular}
\end{center}
\end{table}

\section{Discussion}

For the adopted temperature, the mass of the primary component $M_1 = 0.88$ \ms\ was accepted. Then the semi-major axis $a = 3.1 $ \rs\ was fixed at an appropriate value for the primary mass to be equal to a typical mass of a particular spectral type \citep{2013ApJS..208....9P}. 
With this approach, we were able to estimate the preliminary masses, in addition to the radii and luminosities of both components in absolute units (see Table~\ref{t5}).

No significant flare-like event was recorded during our photometric monitoring, as well as on the precise \T\ light curves.
On the other hand, the precise \T\ light curves of BB~Per shows slowly evolving starspot structures that also affect the estimation of precise \oc\ timings.
Although the resulting colatitude of the dark structure remains practically the same, the value of its longitude changed by 230 degrees in two years and its radius decreased significantly, see Table~\ref{t8}.

The physical parameters of the binary suggest that the times of the tidal synchronization and orbital circularization are very short. These times can be estimated as \citep{2001icbs.book.....H}
\begin{equation}
t_{\rm sync} \simeq 10^4 \, \left[ \frac{1+q}{2q} \right]^2 P^4,  
\end{equation}

\begin{equation}
t_{\rm circ} \simeq 10^6 \, q^{-1} \left[ \frac{1+q}{2} \right]^{5/3} P^{16/3}, 
\end{equation}

\noindent
where both times are in years, $q = M_2/M_1$ is the mass ratio and $P$ is the orbital period in days. For BB~Per these times are
$t_{\rm sync} \simeq 10^3$~yr and
$t_{\rm circ} \simeq 10^5$~yr. 
These times are very short compared to typical ages of late-type binaries
and justify our assumptions about the orbital circularization and synchronicity.

For BB~Per, we also used the publicly available Eclipsing Time Variation calculator
\footnote{Applegate calculator: \url{http://theory-starformation-group.cl/applegate/} } based on the two-zone model by \cite{2016A&A...587A..34V}. For the updated parameters in Table~\ref{t5}, we find that the energy required to drive the Applegate mechanism
is approximately $10^5$ times the available energy in the magnetically active secondary (solution for the finite-shell two-zone model). This mechanism cannot contribute significantly to the observed period changes in this system.

One has to take into consideration that such a small apparent variation in orbital period on a short time scale, frequently attributed to the light-time effect caused by an unseen third body, could be simply a result of the dark-spot evolution or its movement on the surface of binary components.

\section{Conclusions}

\begin{table}
\begin{center}
\caption{Absolute parameters of BB~Per, based on \T\ photometry
and adopted temperatures.}
\smallskip
\label{t5}
\begin{tabular}{ccc}
\hline\hline\noalign{\smallskip}
Parameter         &  Primary     & Secondary  \\
\noalign{\smallskip}\hline\noalign{\smallskip}
Mass [\ms]         &    0.88     &   0.80(6)    \\
Radius [\rs]       &    1.03(2)  &   0.97(2)    \\
Luminosity [L$_{\odot}$]  &  0.696(15)  &  0.510(15)      \\
$M_{\rm bol}$ [mag]&    5.03(2)    &   5.25(2)    \\
$a$ [\rs]          &  \multicolumn{2}{c}{3.1 (fixed)} \\     
\noalign{\smallskip}\hline
\end{tabular}
\end{center}
\end{table}

A study of late-type and low-mass binaries provides us with important information about the most common stars in our Galaxy. 
The new photometric observations of BB~Per as well as \T\ data were used to investigate its light curve and determine the photometric parameters of its components.
The eclipsing system BB~Per contains two similar K-type stars, the more massive and warmer primary component probably with $M_{\rm 1} \simeq 0.88$ \ms\ and the less massive secondary component $M_{\rm 2} = 0.80 \pm 0.06$~\ms. 
Their effective temperatures are $T_1$ = 5300~K and $T_2 = 5050 \pm 50$~K. 
The system inclination is $i=83.6^\circ$.
A dark spot on the surface of the secondary component was revealed, which changed over time. Concerning surface activity, the identification of BB~Per as a faint X-ray source ROSAT 2RXP J035525.1 +313048 also supports our findings \citep{2007A&A...467..785N}.

The current \oc\ diagram based primarily on our new mid-eclipse times and \T\ data shows a periodic change probably caused by a third body orbiting the eclipsing pair with a period of 22~yr. 
The length of this period is strongly affected by the inaccuracy of the first two mid-eclipse times obtained by \cite{2006IBVS.5674....1O} and given in the VarAstro database. 
We can conclude that BB~Per is a nearby low-mass, active, and detached eclipsing binary.
New high-accuracy timings of this eclipsing binary are necessary
to investigate the parameters derived in this paper. 
It is also highly desirable to obtain new spectroscopic observations to obtain the radial-velocity curves and to derive accurate masses for this interesting late-type system. 

\section*{Data Availability}

The \T\ and {\sc Gaia} data underlying this paper are publicly available, and
the other photometric data are available on reasonable request from the authors.

\section*{Acknowledgments}

The authors thank Jan Vraštil, Charles University Prague, 
Zbyněk Henzl, Variable Star and Exoplanet Section, the Czech Astronomical Society, for their kind assistance with photometric observation. 
MW, PZ, JK and JM was supported by the project {\sc Cooperatio-Physics} of Charles University in Prague.
MM was supported by grants from the Ministry of Education of the Czech Republic LM2023032 and LM2023047. 
The research of JM was supported by the Czech Science Foundation (GACR) project
No. 24-10608O.
This publication was produced within the institutional support framework
for the development of the research organization of Masaryk University in Brno.
HK was supported by the RVO 67985815 project.
We thank Eda Sonbas and Adiyaman University Astrophysics Application and Research Center for their support in the acquisition of data with the ADYU60 telescope.
This paper includes data collected by the \T\ mission. 
Funding for the \T\ mission is provided by the NASA's Science Mission Directorate.
This work has made use of data from the European Space Agency (ESA) mission 
{\sc Gaia}~\footnote{\url{https://www.cosmos.esa.int/gaia}}, processed 
by the {\sc Gaia} Data Processing and Analysis Consortium 
(DPAC)~\footnote{\url{https://www.cosmos.esa.int/web/gaia/dpac/consortium}}. 
Funding for the DPAC has been provided by national
institutions, in particular the institutions participating in the Gaia Multilateral Agreement. 
The following Internet-based resources were used in research
for this paper: the SIMBAD and VIZIER database operated at CDS, Strasbourg, France, the NASA's Astrophysics Data System Bibliographic Services, the International Variable Star Index (VSX) database, operated at AAVSO, Cambridge, Massachusetts, USA,  the British Astronomical Association, Photometry Database, and the Czech Astronomical Society B.R.N.O. photometry database of eclipsing binary stars minima and associated O-C Gateway database of minima timings.
This research is part of an ongoing collaboration between professional astronomers, and the Czech Astronomical Society, Variable Star and Exoplanet Section. 


\printcredits

\bibliographystyle{cas-model2-names}

\bibliography{cas-refs}

\end{document}